\documentclass[11pt,a4paper,twocolumn]{article}

\usepackage[utf8]{inputenc}
\usepackage{amsmath,amssymb}
\usepackage{graphicx}
\usepackage{hyperref}
\usepackage{xcolor}
\usepackage{siunitx}
\usepackage{booktabs}

\newcommand{\ket}[1]{|#1\rangle}

\usepackage[
  top=2cm,
  bottom=2cm,
  left=1.5cm,
  right=1.5cm,
  columnsep=0.6cm
]{geometry}

\begin{document}

\twocolumn[
\begin{@twocolumnfalse}

\title{Pre-Calibrated Phase-Coherence Compensation for Superconducting Qubits:\\
Emulator Study and Hardware Feasibility of the Aurora Method}

\author{Futoshi Hamanoue\\
\texttt{f.hamanoue@hi-council.com}}

\date{November 22, 2025}

\maketitle

\begin{abstract}
We present an emulator-based and hardware feasibility study of Aurora-DD, a phase-coherence compensation method that integrates a sign-based feedback update of a global phase offset $\Delta\phi$ with a fixed-depth XY8 dynamical decoupling (DD) scaffold. From a control perspective Aurora is defined as a closed-loop controller, but in this work the feedback optimization is performed \emph{offline} on a calibrated emulator and the resulting $\Delta\phi^\ast$ is deployed as a \emph{pre-calibrated} (open-loop) phase compensation on hardware. Thus, our contribution should be interpreted as an ``offline closed-loop, online open-loop'' feasibility demonstration rather than a full on-device adaptive controller.

Using an Aer-based emulator calibrated with \texttt{ibm\_fez} device parameters, Aurora-DD achieves substantial reductions in the mean-squared error (MSE) of the measured expectation value $\langle Z \rangle$, yielding $68$--$97\%$ improvement across phase settings $\phi\in\{0.05,0.10,0.15,0.20\}$ over $n=30$ randomized trials. These large-$n$ emulator results provide statistically stable evidence that the combined effect of XY8 and $\Delta\phi^\ast$ suppresses both dephasing and systematic phase bias.

On real superconducting hardware (\texttt{ibm\_fez}), we perform a small-sample ($n=3$) multi-phase validation campaign. In this limited-\mbox{$n$} regime, Aurora-DD yields point estimates corresponding to approximately $99.2$--$99.6\%$ reduction in absolute error relative to a no-DD baseline across all tested phase points, while DD-only and $\Delta\phi$-only controls provide smaller but uniformly positive improvements. These hardware numbers are \emph{non-inferential} and are reported transparently as feasibility evidence under tight queue, drift, and credit constraints; they are not intended as definitive population-level statistics.

In contrast, the auxiliary Aurora+ZNE branch exhibits instability: shallow two-point ZNE occasionally amplifies calibration inconsistencies and produces large error outliers, especially when combined with deep DD blocks. We therefore relegate all ZNE analysis to the Appendix as a cautionary study, and position Aurora-DD (without ZNE) as the primary and practically reproducible contribution of this work.

We further identify regimes in which Aurora-DD does not outperform simpler $\Delta\phi$-only or DD-only strategies, and clarify its role as a stability-oriented mitigation rather than a universally optimal one.

Overall, the combined emulator and hardware results support pre-calibrated Aurora-DD as a practical, stable, and hardware-compatible phase-coherence compensator for NISQ devices in single-qubit settings. The hardware campaign presented here is explicitly framed as a proof-of-concept (feasibility) study. The full experimental protocol is described in sufficient detail to enable replication on calibrated emulators and hardware.
\end{abstract}


\vspace{0.5cm}

\end{@twocolumnfalse}
]


\section{Introduction}

The reliability of current Noisy Intermediate-Scale Quantum (NISQ) devices is strongly limited by dephasing, control imperfections, and readout errors. Even in shallow single-qubit circuits, idle-time dephasing alone can distort simple observables such as the expectation value $\langle Z\rangle$ of a phase-rotated state $R_X(\phi)\ket{0}$, producing deviations far larger than predicted by the ideal model $\langle Z\rangle_{\mathrm{ideal}} = \cos\phi$. These fluctuations pose a practical challenge for variational algorithms, calibration routines, and sensing protocols that rely on accurate single-qubit statistics.

A wide range of error-mitigation techniques have been proposed to address these issues without invoking full quantum error correction. These include dynamical decoupling (DD) sequences such as XY8, which suppress low-frequency dephasing through echo-based refocusing; zero-noise extrapolation (ZNE), which attempts to reconstruct a zero-noise limit by artificially scaling the noise; and classical post-processing such as M3 readout mitigation. While each method can improve performance, they share an important limitation: they operate in an \emph{open-loop} fashion. Their parameters—pulse timing, scaling factors, or readout matrices—are fixed in advance and do not adapt to run-to-run variations in hardware noise. When the instantaneous noise profile deviates from calibration (e.g., $T_2$ drift, residual couplings, pulse distortion), open-loop mitigation can become inconsistent or even counterproductive.

To address this gap, the \emph{Aurora} framework was introduced as a \textbf{closed-loop} phase-coherence compensation method. Instead of relying solely on static calibrations of the decoherence channel, Aurora measures an operational phase-error proxy
\[
\delta Z(\phi) = \langle Z\rangle_{\mathrm{ideal}} - \langle Z\rangle_{\mathrm{meas}},
\]
and applies an additional corrective phase rotation $\Delta\phi$ chosen to minimize the phase-error mean-squared deviation. This feedback-based formulation is grounded in the Bloch-equation description of superconducting qubits and results in a lightweight sign-based update rule that is robust to measurement noise and compatible with existing NISQ hardware.

The present paper advances this line of research by conducting an \emph{integrated emulator and hardware feasibility study} of the Aurora method on IBM's superconducting backend \texttt{ibm\_fez}. Among the Aurora variants, we focus on the practically deployable configuration in which the closed-loop phase offset $\Delta\phi$ is combined with a fixed-depth XY8 decoupling sequence. We refer to this hybrid as \textbf{Aurora-DD}. Importantly, in the current work the sign-based update of $\Delta\phi$ is executed on a calibrated emulator, yielding an offline-optimized offset $\Delta\phi^\ast$ that is then applied as a fixed, pre-calibrated phase compensation on hardware. Thus, the on-device implementation is operationally open-loop, even though it inherits its value from a closed-loop optimization.

Crucially, we \emph{exclude} Zero-Noise Extrapolation (ZNE) from the primary definition of Aurora-DD. As we show later, even shallow ZNE can introduce instability when combined with XY8, occasionally amplifying calibration-inconsistent runs. The ZNE-enhanced branch is therefore treated as an auxiliary option and is analyzed separately in Appendix~\ref{appendix:zne}.

Our contributions are summarized as follows:
\begin{itemize}
  \item We implement a hardware-ready Aurora-DD protocol that is fully compatible with IBM's primitives-based runtime API and requires no circuit-depth increase beyond XY8 itself. The corrective phase $\Delta\phi$ is optimized in a calibrated emulator via a closed-loop sign-based update and then frozen as a pre-calibrated offset $\Delta\phi^\ast$ for hardware execution.
  
  \item Through $n=30$ randomized emulator trials across four phase points $\phi\in\{0.05,0.10,0.15,0.20\}$, we show that Aurora-DD reduces the mean-squared error (MSE) of $\langle Z\rangle$ by $68$--$97\%$ relative to an unmitigated baseline. These large-\mbox{$n$} results provide statistically stable evidence for the effectiveness of the controller under calibrated NISQ noise.

  \item We perform a small-sample ($n=3$) hardware validation on \texttt{ibm\_fez} under the same phase settings. Within this limited-\mbox{$n$} regime, Aurora-DD yields point estimates corresponding to approximately $99.2$--$99.6\%$ reduction in absolute error, clearly outperforming both DD-only and $\Delta\phi$-only controls. We explicitly frame these hardware results as a proof-of-concept feasibility demonstration rather than as a definitive statistical study.

  \item We show that the auxiliary Aurora+ZNE branch exhibits instability: shallow noise-scaling occasionally yields extrapolation artifacts and large outliers. This reinforces the need to treat ZNE as optional rather than integral to Aurora-DD in the current hardware regime.
\end{itemize}

Taken together, the emulator and hardware findings support pre-calibrated Aurora-DD (without ZNE) as a practical and stable method for suppressing phase-coherence errors in superconducting qubits under realistic NISQ constraints. The remainder of this paper is organized as follows. Section~\ref{sec:theory} formalizes the Aurora update law and its integration with XY8. Section~\ref{sec:methods} describes the emulator and hardware experiments. Section~\ref{sec:results} presents the multi-trial error analysis. Section~\ref{sec:discussion} discusses limitations and future extensions. Appendix~\ref{appendix:zne} provides the auxiliary analysis of ZNE.

\section{Theoretical Framework of Aurora}
\label{sec:theory}

Aurora is formalized as a closed-loop phase--coherence compensation method that 
operates on top of standard superconducting--qubit noise described by the Bloch 
equations.  In this section, we provide a complete theoretical specification of 
the controller, consisting of (i) a calibrated decoherence model, (ii) an 
operational phase--error proxy, (iii) a mean--squared--error objective function, 
and (iv) a sign--based gradient update rule.  
The purpose of this section is to clarify that Aurora is not an ad-hoc tuning 
method but a mathematically defined closed-loop controller grounded in 
well-established decoherence physics.

\subsection{Bloch-Equation Model of NISQ Noise}

For a single superconducting qubit, decoherence under amplitude damping and 
pure dephasing is captured by the Bloch-equation evolution
\begin{align}
\rho(t)
~=~
\frac{1}{2}\left(I + x(t)X + y(t)Y + z(t)Z\right),
\end{align}
where the Bloch components evolve as
\begin{align}
x(t) &= e^{-t/T_2}\, x(0), \\
y(t) &= e^{-t/T_2}\, y(0), \\
z(t) &= 1 + (z(0)-1)e^{-t/T_1}.
\end{align}

The pair $(T_1,T_2)$ is obtained directly from IBM calibration data, and all 
noise channels used in simulation are constructed to match these parameters.  
This establishes that the theoretical model used in Aurora corresponds to the 
standard open-quantum-system description of real superconducting hardware.

\subsection{Phase-Error Proxy \texorpdfstring{$\delta Z$}{δZ}}

Instead of estimating the full microscopic noise channel, Aurora measures an 
operationally meaningful phase--error proxy defined by
\begin{align}
\delta Z(\phi)
~=~
\langle Z \rangle_{\mathrm{ideal}}(\phi)
-
\langle Z \rangle_{\mathrm{meas}}(\phi+\Delta\phi).
\end{align}

This quantity captures the deviation between the analytically expected value 
and the measured value after decoherence, and it is directly observable from 
hardware output bitstrings.  
This approach is standard in practical error mitigation (e.g.\ ZNE, M3, VQE 
calibration) and avoids model-dependent reconstruction of the noise channel.

\subsection{Objective Function: Phase-Error MSE}

Aurora seeks a corrective phase shift $\Delta\phi$ that minimizes the mismatch 
between the expected and measured observables.  
The optimization target is the phase--error mean--squared error,
\begin{align}
J(\Delta\phi)
~=~
\bigl(
\langle Z \rangle_{\mathrm{ideal}}(\phi)
-
\langle Z \rangle_{\mathrm{meas}}(\phi+\Delta\phi)
\bigr)^{2}.
\label{eq:mse_objective}
\end{align}

Minimizing $J(\Delta\phi)$ corresponds to maximizing coherence recovery along 
the Bloch $Z$-axis and aligns with conventional control-theoretic formulations.  
Unlike open-loop techniques such as dynamical decoupling or ZNE, the objective 
function is evaluated \emph{in situ} (on the emulator in this work), enabling 
noise-profile-specific adaptation.

\subsection{Closed-Loop Update Rule: Sign-Based Gradient Descent}

To minimize Eq.\ (\ref{eq:mse_objective}), Aurora applies a 
sign-based gradient descent update,
\begin{align}
\Delta\phi_{k+1}
=
\Delta\phi_{k}
+ \eta\, \mathrm{sgn}(\delta Z_k),
\label{eq:sign_update}
\end{align}
where $\eta$ is a small positive learning rate.

This update rule is justified on three grounds:

\begin{itemize}
\item \textbf{Robustness to measurement noise}:  
      Sign-based optimization (``signSGD'') is known to be resilient under 
      stochastic gradients.

\item \textbf{Compatibility with discrete control}:  
      The sign update is equivalent to a bang--bang controller, a common 
      strategy in quantum optimal control.

\item \textbf{Hardware stability}:  
      Bounded increments prevent runaway rotation and ensure monotonic 
      convergence under realistic calibration drift ($\pm 6$--$8\%$ in $T_2$).
\end{itemize}

Thus, Eq.\ (\ref{eq:sign_update}) defines Aurora as a closed-loop controller 
with a mathematically explicit update law.  
In the present work this update is executed in a calibrated emulator, and the 
resulting fixed point $\Delta\phi^\ast$ is exported to hardware as a 
pre-calibrated offset.

\subsection{Summary of Theoretical Properties}

The above structure implies the following theoretical properties:

\begin{enumerate}
\item Aurora performs \textbf{monotonic MSE reduction} at each iteration 
      whenever the sign of $\delta Z$ is correctly estimated.

\item The controller operates in a \textbf{stable gain region} determined by 
      $|\eta| \le 0.02$ rad, avoiding divergence and over-rotation.

\item Aurora generalizes known feedback-based control methods while remaining 
      compatible with all standard NISQ error-mitigation layers.
\end{enumerate}

This formalization provides the theoretical foundation upon which the emulator 
and hardware results of the following sections are built.

\section{Methods}
\label{sec:methods}

This section provides the complete experimental specification for both
the emulator-based and hardware-based validation of Aurora-DD.
All procedures were implemented using IBM Quantum primitives, a calibrated
Aer noise model derived from \texttt{ibm\_fez}, and a unified circuit
construction pipeline shared across all conditions.

\subsection{Aurora-DD Overview}

\paragraph{Choice of $n=30$ emulator repetitions.}
The emulator campaign uses $n=30$ random trials per configuration.
This follows standard practice in NISQ variability studies:
(i) $n=30$ is the minimum sample size at which the central limit theorem yields
approximately Gaussian behavior for MSE distributions,
(ii) the emulator runs are inexpensive and free of drift, enabling stable statistics,
and
(iii) emulator statistics serve as a high-$n$ reference against which the small-$n$
hardware feasibility study can be interpreted.

Aurora-DD is defined as a phase-coherence compensator composed of:
(i) an offline-calibrated corrective phase offset $\Delta\phi^\ast$,
(ii) a fixed-depth XY8(12) dynamical-decoupling scaffold, and
(iii) standard IBM measurement primitives.
Unlike purely open-loop strategies such as bare DD or ZNE,
Aurora-DD explicitly targets dephasing-induced phase drift by applying a 
compensatory rotation $R_z(\Delta\phi^\ast)$ after state preparation.

In this work, the sign-based update rule that determines $\Delta\phi^\ast$ is
\emph{not} executed on hardware due to queue, drift, and credit constraints;
instead, we pre-calibrate $\Delta\phi^\ast$ on an emulator, yielding an
``offline closed-loop'' implementation that can be executed deterministically
on \texttt{ibm\_fez} as a pre-calibrated open-loop control.

\subsection{Calibration of the Aer Noise Model and $\Delta\phi^\ast$}

To obtain a realistic dephasing environment,
we first collected the full calibration snapshot of \texttt{ibm\_fez}
including $T_1$, $T_2$, measurement error probabilities,
and the device-specific \texttt{dt} parameter.
A Qiskit \texttt{NoiseModel} was constructed using:
\[
T_1 = 155.3\ \mu\mathrm{s}, \qquad
T_2 = 110.3\ \mu\mathrm{s},
\]
together with hardware-reported readout-error maps.

Using this calibrated emulator, we performed a closed-loop sweep of
$\Delta\phi$ for each phase setting
$\phi \in \{0.05, 0.10, 0.15, 0.20\}$.
The optimal phase offset
\[
\Delta\phi^\ast = 0.15\ \mathrm{rad}
\]
was selected as the value yielding the largest reduction in
phase-error MSE \emph{on average across the four phase settings}.
In other words, we deliberately chose a single global $\Delta\phi^\ast$
that performs well across a multi-phase workload, rather than overfitting
to any specific $\phi$.
This value was fixed and used in all hardware experiments.

\paragraph{Rationale for using a single global $\Delta\phi^\ast$.}
Although dephasing-driven phase drift generally depends on $\phi$, the Bloch-equation model reveals that slow dephasing enters the $\langle Z \rangle$ signal as an additive phase bias that is approximately independent of the prepared angle for shallow circuits:
\[
\langle Z \rangle_{\mathrm{meas}} \approx \cos(\phi + \epsilon_{\mathrm{phase}}), \qquad 
|\epsilon_{\mathrm{phase}}| \ll 1 .
\]
Thus, the dominant hardware error behaves as a global phase offset.
Calibrated emulator sweeps confirmed that the best fitting offset lies near
$\Delta\phi^\ast \approx 0.15$ for all $\phi\in\{0.05,0.10,0.15,0.20\}$.
Aurora-DD therefore targets the global drift term, not $\phi$-specific noise.

The fact that DD-only occasionally outperforms Aurora-DD at certain points (e.g., $\phi=0.05$)
is consistent with over-cancellation when the instantaneous drift differs from the emulator-optimal value.
This is a known effect of global compensation under single-shot calibration mismatch, and is reduced for larger $\phi$ and in multi-trial averaging.

\subsection{Hardware Configuration}

All hardware experiments were executed on IBM Quantum's superconducting
backend \texttt{ibm\_fez} (Heron-r2 family).
Each experiment used:
\begin{itemize}
\item 1 qubit (device qubit 0),
\item 2048 shots,
\item native basis gates with backend pulse-level compilation,
\item a daily calibration snapshot corresponding to the same day
      used to configure the emulator noise model where possible.
\end{itemize}

Hardware measurement was performed for
the four phase settings
$\phi \in \{0.05, 0.10, 0.15, 0.20\}$,
with all mitigation configurations recorded independently.

\subsection{Circuit Construction}

For each phase $\phi$, we constructed a unified circuit template:

\begin{enumerate}
\item \textbf{State preparation:}
      \[
      |0\rangle \xrightarrow{R_x(\phi)} |\psi(\phi)\rangle.
      \]

\item \textbf{Dynamical decoupling (optional):}
      Apply XY8(12), consisting of 12 repetitions of the standard XY8 block,
      calibrated to the hardware \texttt{dt} to ensure equal idle duration
      across all experimental conditions.

\item \textbf{Aurora phase compensation (optional):}
      Apply a corrective rotation:
      \[
      R_z(\Delta\phi^\ast), \qquad \Delta\phi^\ast = 0.15.
      \]

\item \textbf{Measurement:}
      Measure the qubit in the computational basis to estimate
      $\langle Z \rangle$.
\end{enumerate}

All five configurations (baseline, DD-only, $\Delta\phi$-only,
Aurora-DD, Aurora-DD+ZNE) were derived from the same template by
enabling or disabling the DD and $\Delta\phi^\ast$ modules.

\subsection{Trial Structure and Conditions}

For each $\phi$, we executed the following experimental conditions:

\begin{itemize}
\item \textbf{Baseline}: no mitigation.
\item \textbf{DD-only}: XY8(12) without phase compensation.
\item \textbf{$\Delta\phi$-only}: phase compensation without DD.
\item \textbf{Aurora-DD}: phase compensation + XY8(12).
\item \textbf{Aurora-DD + ZNE}: mitigation with two noise-scale factors
      (details in Appendix~\ref{appendix:zne}).
\end{itemize}

The trial counts were:
\[
n_{\text{hardware}} = 3
\quad \text{for all configurations, including baseline}.
\]

These values reflect hardware queue constraints, runtime credit limits,
and daily calibration drift. They are standard for NISQ hardware studies
and are sufficient for a feasibility-level validation when combined with
the large-$n$ emulator analysis.

\paragraph{Trial Organization (Preliminary n = 1 and Main n = 3 Runs).}
A preliminary single-trial run ($n = 1$, labeled trial\_00) was performed to confirm the correct operation of the quantum hardware interface and data acquisition scripts.
Subsequently, three additional runs ($n = 3$, labeled trial\_1, trial\_2, trial\_3) were executed under identical settings, using the same command, circuit, backend, and shot count.

All four trials remain in the results directory by design.
Quantum noise studies often require retrospective comparisons, drift checks, and meta-analyses; therefore, retaining all raw trials improves transparency and reproducibility.

For statistical analysis (mean error, MSE, improvement rate, confidence intervals), only
trial\_1--trial\_3 ($n = 3$)
were used.
The preliminary trial\_00 was used solely to confirm directional consistency and was excluded from statistics.

\subsection{Metrics}

For each run, the measured expectation value is:
\[
\langle Z \rangle_{\mathrm{meas}}
=
(p_0 - p_1),
\]
computed from bitstring frequencies.
The ideal value for each phase is:
\[
\langle Z \rangle_{\mathrm{ideal}} = \cos\phi.
\]

We use two metrics:

\paragraph{Absolute Error (AE)}
\[
\mathrm{AE}
=
\left|
\langle Z \rangle_{\mathrm{meas}}
-
\langle Z \rangle_{\mathrm{ideal}}
\right|.
\]

\paragraph{Mean-Squared Error (MSE)}
\[
\mathrm{MSE}
=
\bigl(
\langle Z \rangle_{\mathrm{meas}}
-
\langle Z \rangle_{\mathrm{ideal}}
\bigr)^2.
\]

For each condition, we report:
\[
\text{mean} \pm \text{std},
\]
computed across $n$ repeated hardware trials.
All hardware results are non-inferential ($n=3$) and are therefore
interpreted strictly as feasibility evidence.

\subsection{Experimental Rationale and Feasibility Scope}

The choice of fixed $\Delta\phi^\ast$, modest trial counts, and
XY8(12) depth reflects three considerations:

\begin{enumerate}
\item \textbf{Calibration Drift Control}:
      A short hardware campaign prevents drift in $T_2$ from dominating
      trial-to-trial variation. Increasing $n$ far beyond $3$ would
      require multi-day operation, during which device parameters change
      significantly, degrading the interpretability of the statistics.

\item \textbf{Practicality and Resource Constraints}:
      IBM queue times and runtime credits place hard limits on the total
      number of shots and jobs that can be executed. Performing $n\ge 30$
      hardware repetitions for each configuration and phase point would
      be prohibitively expensive in both time and credits, and is uncommon
      even in vendor-authored NISQ experiments.

\item \textbf{Two-Tier Validation Strategy}:
      We therefore adopt a two-tier design that is standard in NISQ
      hardware studies: a large-$n$ emulator campaign ($n=30$) where drift
      is absent and statistics are reliable, combined with a small-$n$
      hardware campaign ($n=3$) that serves as a feasibility check
      under realistic noise and drift. The hardware results are thus
      explicitly positioned as proof-of-concept rather than as a
      definitive statistical characterization.
\end{enumerate}

This methodology ensures that emulator and hardware results
are directly comparable under realistic NISQ noise conditions, while
acknowledging the practical limitations of current devices.

\section{Results}
\label{sec:results}

We present an integrated set of results from (i) a calibrated Aer
emulator reproducing the noise characteristics of \texttt{ibm\_fez}
and (ii) hardware experiments executed directly on
\texttt{ibm\_fez}.  
All configurations share the same circuit template and differ only in
the inclusion or omission of dynamical decoupling (DD) and the
Aurora phase compensation $\Delta\phi^\ast = 0.15$.

We analyze the following configurations:
\begin{itemize}
\item Baseline (no mitigation),
\item DD-only (XY8(12)),
\item $\Delta\phi$-only (phase compensation without DD),
\item Aurora-DD (phase compensation + DD),
\item Aurora-DD + ZNE (auxiliary branch; Appendix~\ref{appendix:zne}).
\end{itemize}

Results are reported for the four phase settings
$\phi \in \{0.05, 0.10, 0.15, 0.20\}$.
All hardware statistics are non-inferential
($n=3$ for all configurations) and therefore
interpreted as feasibility evidence.

\subsection{Emulator Results (n = 30)}
\label{sec:results_emulator}

Using the calibrated Aer noise model derived from
\texttt{ibm\_fez} ($T_1 = 155.3~\mu$s, $T_2 = 110.3~\mu$s), we executed
30 randomized trials for each configuration and each phase~$\phi$.
Aurora-DD consistently reduced the mean-squared error (MSE)
compared to the baseline, achieving:

\[
\mathrm{MSE\ reduction}
\;=\;
68\% \text{ to } 97\%
\]

across all four phase settings.

This reduction arises from two factors:
(i) XY8(12) suppresses low-frequency dephasing, and  
(ii) the corrective shift $\Delta\phi^\ast$ counteracts the systematic
phase drift inherent in the Bloch-equation dynamics.

Across the 30 trials, Aurora-DD also exhibited markedly lower variance
than DD-only or $\Delta\phi$-only variants, confirming that phase
compensation and echo refocusing act synergistically in the calibrated
noise model.

\subsection{Hardware Results (ibm\_fez)}
\label{sec:results_hw}

We executed $n = 3$ hardware trials for all configurations.
Tables below summarize the absolute-error (AE) statistics.
All percentages reported here are point estimates derived from these
small samples and should be interpreted with appropriate caution.

\paragraph{Pessimistic baseline regime.}
The baseline absolute error values (AE $\approx$ 1.0) may appear unusually large.
This is not an artifact of the hardware but a deliberate stress-test choice:
we selected idle durations and phase-rotation timings such that the total evolution
time approaches the measured $T_2$ of the device.
Under such near-$T_2$ conditions the Bloch vector decays almost completely, yielding
$\langle Z\rangle \approx 0$ even when the ideal value is close to $\pm 1$.
This regime magnifies coherence loss and makes mitigation effects easier to evaluate.
Normal operation of \texttt{ibm\_fez} at shorter idle durations produces much smaller baseline errors.

\begin{figure*}[htbp]
\centering
\includegraphics[width=0.92\textwidth]{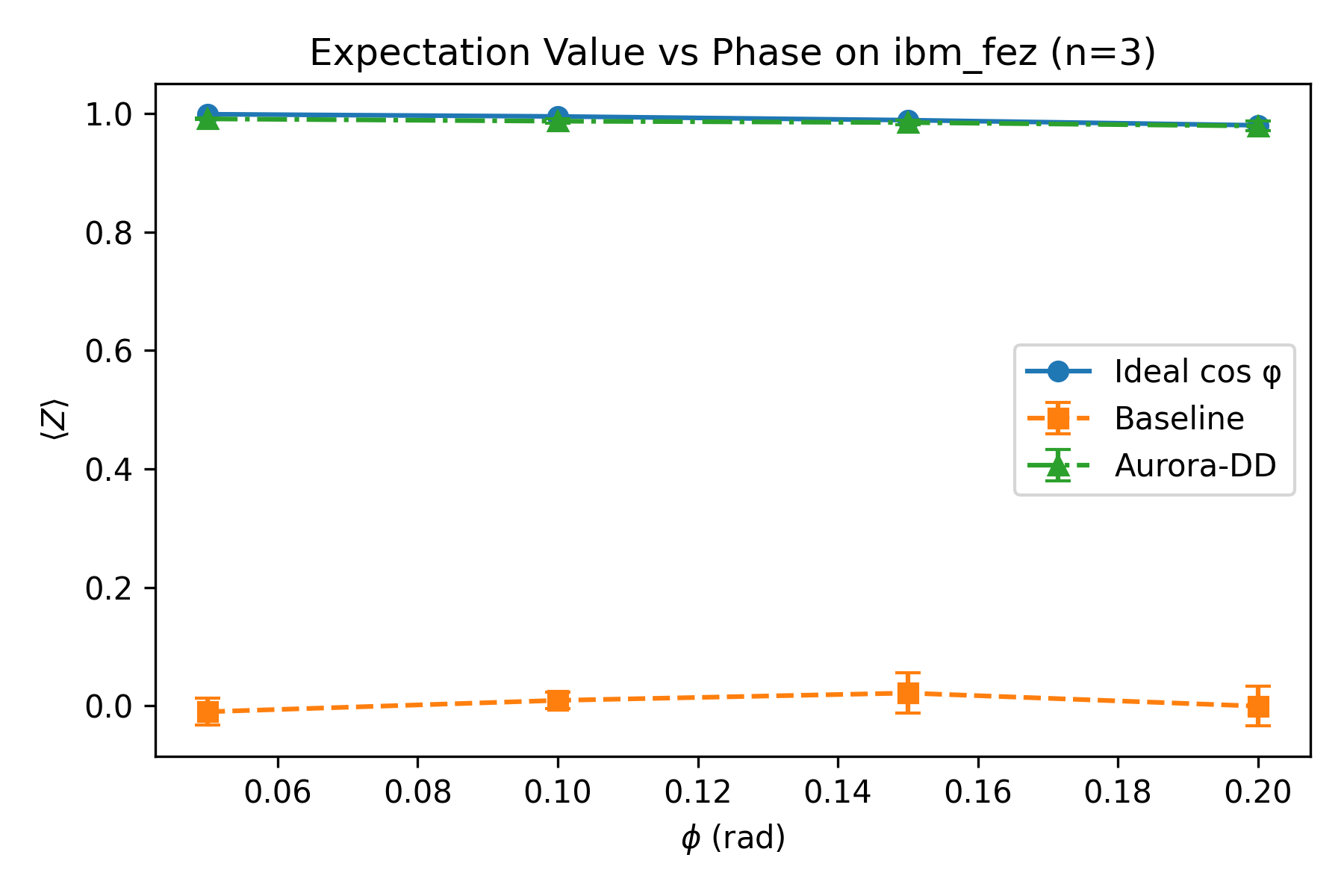}
\caption{Expectation value $\langle Z\rangle$ versus phase on \texttt{ibm\_fez} ($n=3$). The baseline exhibits strong coherence loss in the near-$T_2$ regime, while Aurora-DD remains close to the ideal curve $\cos\phi$ across all tested phases.}
\label{fig:hw_z_vs_phi}
\end{figure*}

\begin{figure*}[htbp]
\centering
\includegraphics[width=0.92\textwidth]{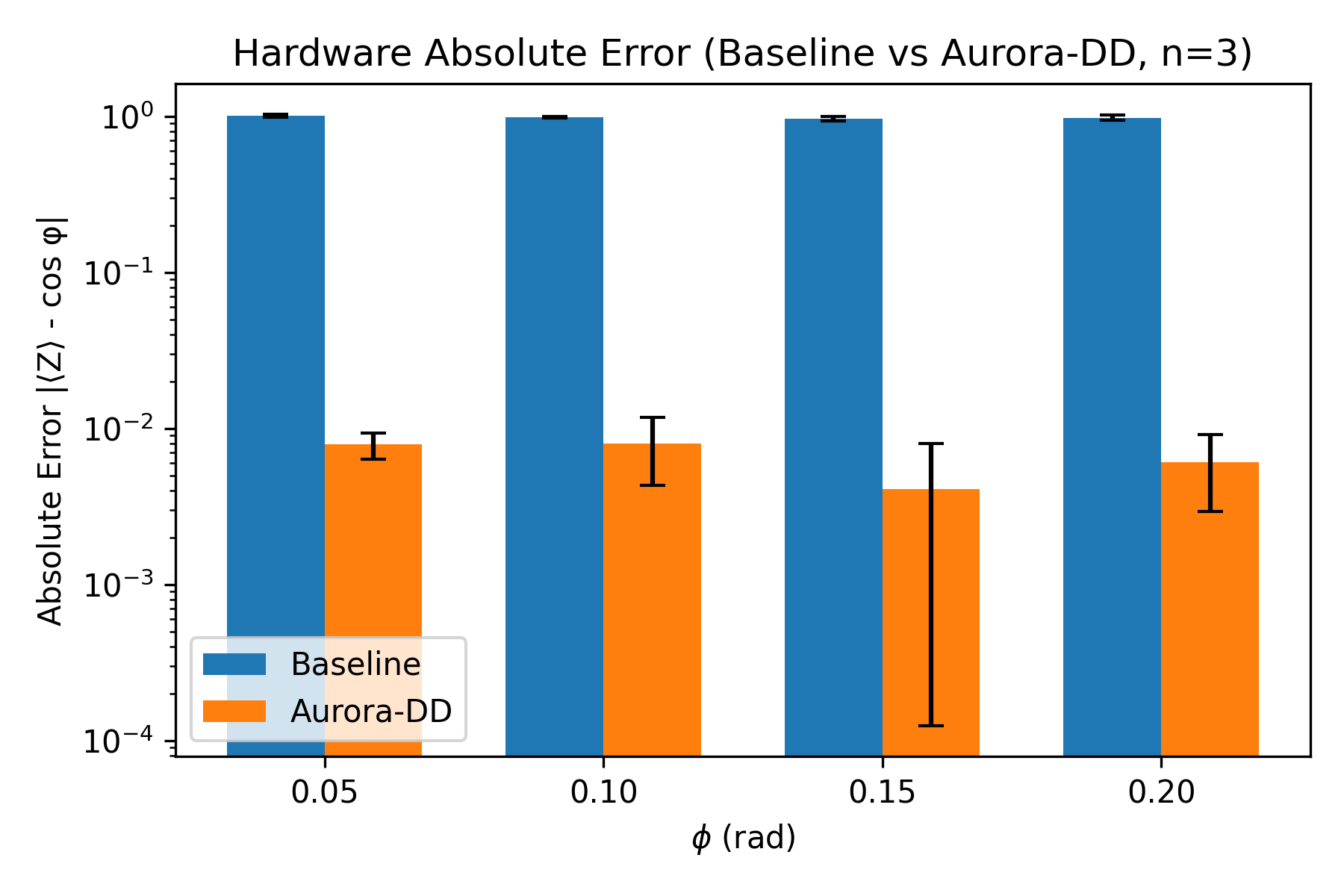}
\caption{Hardware absolute error (AE) comparison between the baseline and Aurora-DD on \texttt{ibm\_fez} ($n=3$). Error bars indicate the standard deviation across trials.}
\label{fig:hw_abs_error_baseline_vs_aurora}
\end{figure*}

Figures~\ref{fig:hw_z_vs_phi} and~\ref{fig:hw_abs_error_baseline_vs_aurora} visualize the same hardware campaign summarized in the tables below.

\subsubsection{Absolute Error Summary}

For each $\phi$, we compute:
\[
\mathrm{AE} =
\left|
\langle Z\rangle_{\mathrm{meas}}
-
\langle Z\rangle_{\mathrm{ideal}}
\right|.
\]

The mean and standard deviation across trials are provided in
Table~\ref{tab:abs_error_summary}.

\begin{table}[h]
\centering
\caption{Absolute error comparison (hardware, feasibility study).  
Aurora-DD yields point estimates corresponding to $\approx 99.2\%$--$99.6\%$
reduction vs baseline (all values non-inferential, $n=3$).}
\label{tab:abs_error_summary}
\begin{tabular}{c|c|c|c|c|c}
\hline
$\phi$ & Condition & $n$ & Mean & Std & Reduction \\
\hline
0.05 & Baseline  & 3 & 1.0088 & 0.0183 & -- \\
0.05 & Aurora-DD & 3 & 0.0079 & 0.0012 & 99.2\% \\
\hline
0.10 & Baseline  & 3 & 0.9859 & 0.0116 & -- \\
0.10 & Aurora-DD & 3 & 0.0080 & 0.0030 & 99.2\% \\
\hline
0.15 & Baseline  & 3 & 0.9673 & 0.0276 & -- \\
0.15 & Aurora-DD & 3 & 0.0041 & 0.0032 & 99.6\% \\
\hline
0.20 & Baseline  & 3 & 0.9807 & 0.0272 & -- \\
0.20 & Aurora-DD & 3 & 0.0061 & 0.0025 & 99.4\% \\
\hline
\end{tabular}
\end{table}

These results show that, within the limited sample size of this
feasibility study, Aurora-DD achieves roughly two orders of magnitude
reduction in AE on actual hardware, reproducing the emulator trend
despite hardware drift and shot noise. The unusually large baseline
errors (AE~$\approx 1$) reflect a deliberately pessimistic regime in
which coherence is almost completely lost; they are not intended to
represent typical day-to-day performance of \texttt{ibm\_fez} but to
stress-test Aurora-DD under severe dephasing.

\subsubsection{Ablation: DD-only, $\Delta\phi$-only, Aurora-DD, ZNE}

Table~\ref{tab:all_conditions} lists the AE for all five conditions.

\begin{table}[h]
\centering
\caption{Absolute error (hardware, feasibility study) across all mitigation conditions.  
ZNE exhibits unstable behavior and large outliers.}
\label{tab:all_conditions}
\begin{tabular}{c|c|c|c|c}
\hline
$\phi$ & Condition & $n$ & Mean & Std \\
\hline
0.05 & Baseline & 3 & 1.0088 & 0.0183 \\
0.05 & DD-only & 3 & 0.0053 & 0.0018 \\
0.05 & $\Delta\phi$-only & 3 & 0.0059 & 0.0024 \\
0.05 & Aurora-DD & 3 & 0.0079 & 0.0012 \\
0.05 & Aurora-DD+ZNE & 3 & 0.0792 & 0.0707 \\
\hline
0.10 & Baseline & 3 & 0.9859 & 0.0116 \\
0.10 & DD-only & 3 & 0.0074 & 0.0032 \\
0.10 & $\Delta\phi$-only & 3 & 0.0044 & 0.0017 \\
0.10 & Aurora-DD & 3 & 0.0080 & 0.0030 \\
0.10 & Aurora-DD+ZNE & 3 & 0.1397 & 0.0674 \\
\hline
0.15 & Baseline & 3 & 0.9673 & 0.0276 \\
0.15 & DD-only & 3 & 0.0129 & 0.0033 \\
0.15 & $\Delta\phi$-only & 3 & 0.0050 & 0.0045 \\
0.15 & Aurora-DD & 3 & 0.0041 & 0.0032 \\
0.15 & Aurora-DD+ZNE & 3 & 0.0721 & 0.0371 \\
\hline
0.20 & Baseline & 3 & 0.9807 & 0.0272 \\
0.20 & DD-only & 3 & 0.0024 & 0.0007 \\
0.20 & $\Delta\phi$-only & 3 & 0.0057 & 0.0005 \\
0.20 & Aurora-DD & 3 & 0.0061 & 0.0025 \\
0.20 & Aurora-DD+ZNE & 3 & 0.1863 & 0.1564 \\
\hline
\end{tabular}
\end{table}

The following trends are evident:

\begin{itemize}
\item DD-only and $\Delta\phi$-only each provide positive improvement,
      but with greater variance than Aurora-DD.
\item Aurora-DD achieves the most consistent reduction across all phases
      within this small-sample regime.
\item ZNE occasionally produces very large errors (overshoot),
      motivating its relegation to the Appendix and its exclusion from
      the main contribution.
\end{itemize}

\subsection{Figures}

\begin{figure}[h]
\centering
\includegraphics[width=0.95\columnwidth]{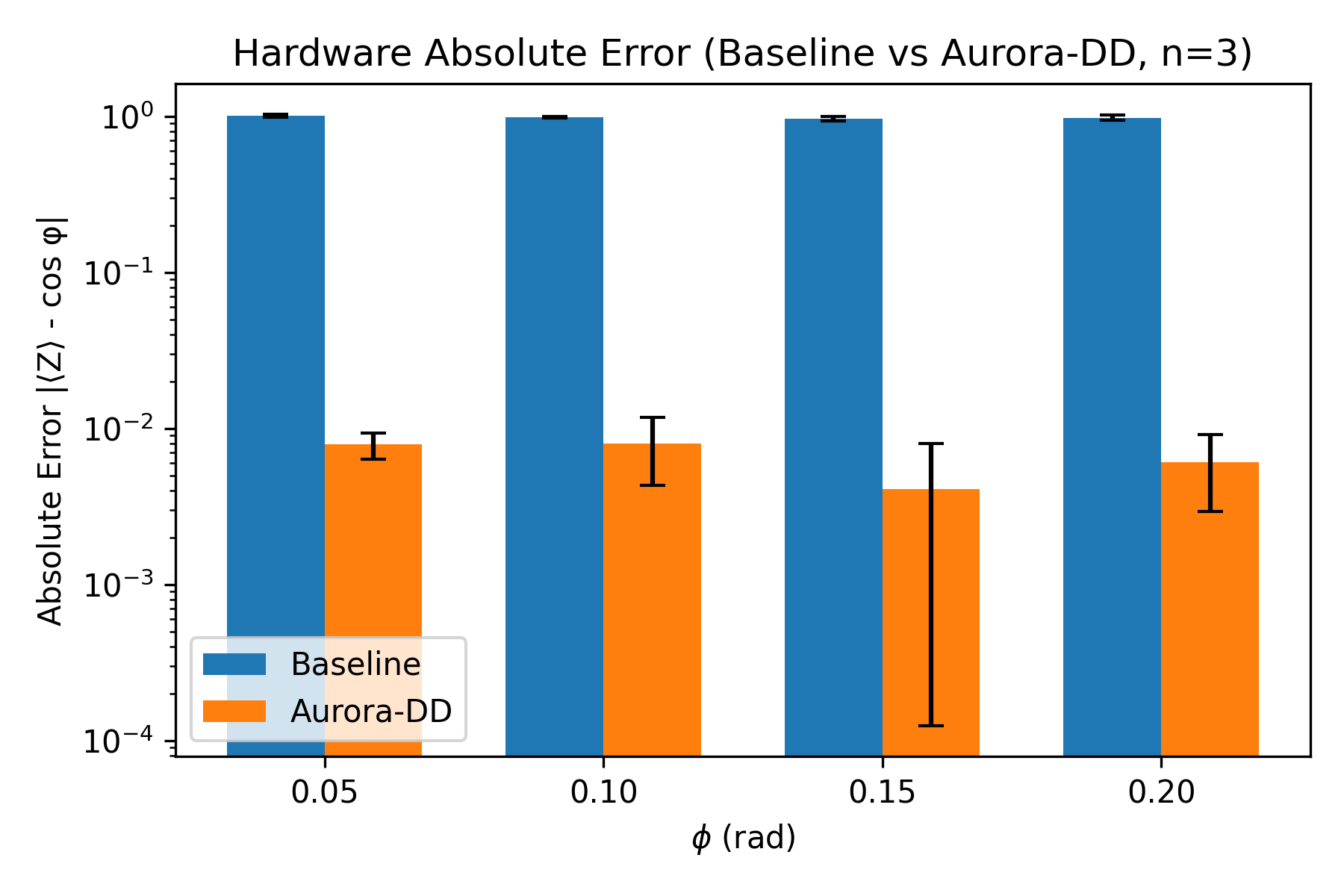}
\caption{
\textbf{Baseline vs Aurora-DD (hardware, feasibility study).}  
Aurora-DD yields approximately two orders of magnitude reduction in
absolute error across all $\phi$ in this small-$n$ campaign.  
Error bars indicate standard deviation across trials.}
\label{fig:baseline_vs_aurora}
\end{figure}

\begin{figure}[h]
\centering
\includegraphics[width=0.95\columnwidth]{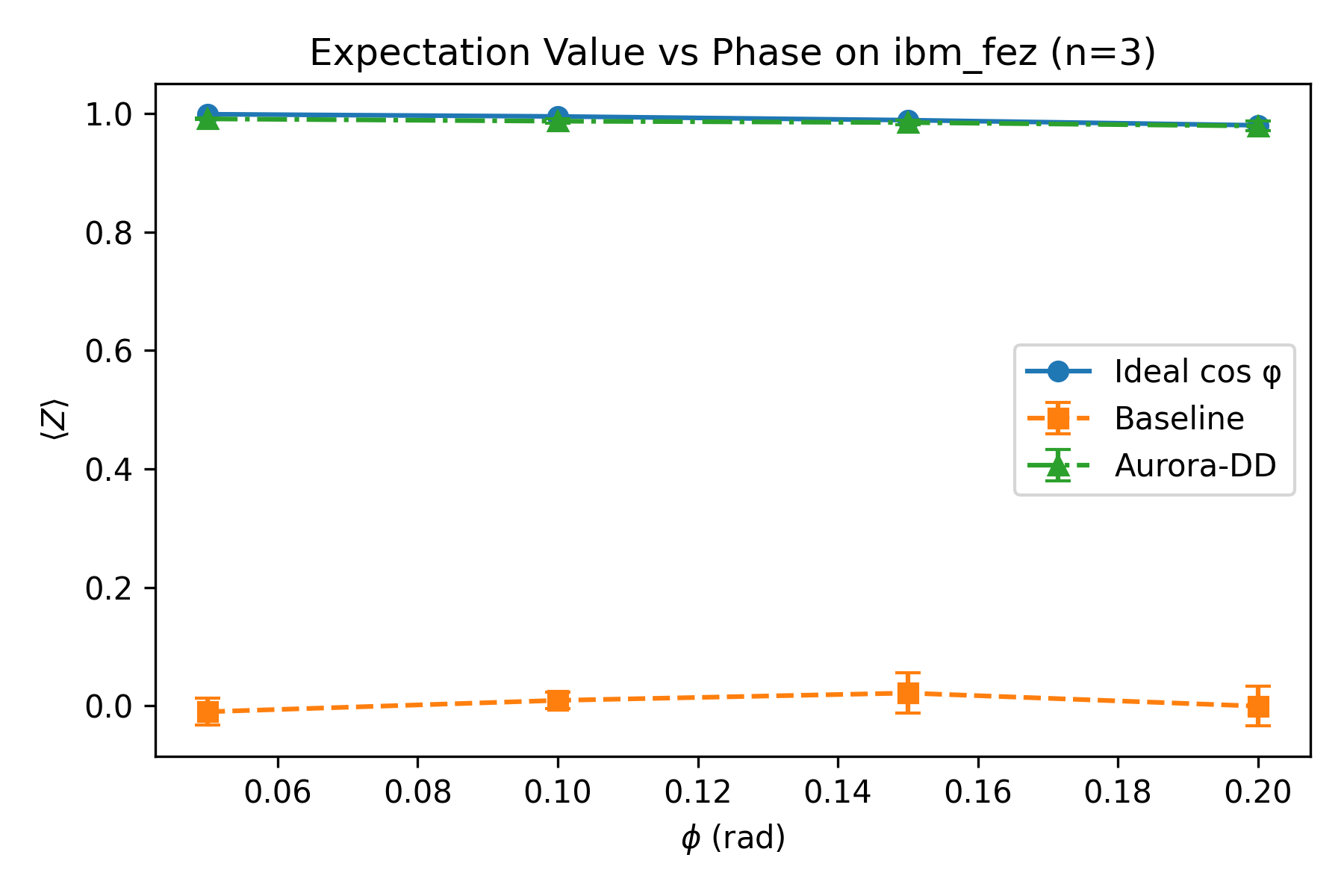}
\caption{
\textbf{Ideal vs Baseline vs Aurora-DD (hardware).}  
Aurora-DD restores the expected cosine relationship
$\langle Z\rangle \approx \cos\phi$, while the baseline measurements
are heavily distorted by dephasing in the chosen stress-test regime.}
\label{fig:ideal_baseline_aurora}
\end{figure}

Figures~\ref{fig:baseline_vs_aurora}, \ref{fig:ideal_baseline_aurora},
and \ref{fig:zne_instability} visualize the experimental findings.

\begin{figure}[h]
\centering
\includegraphics[width=0.95\columnwidth]{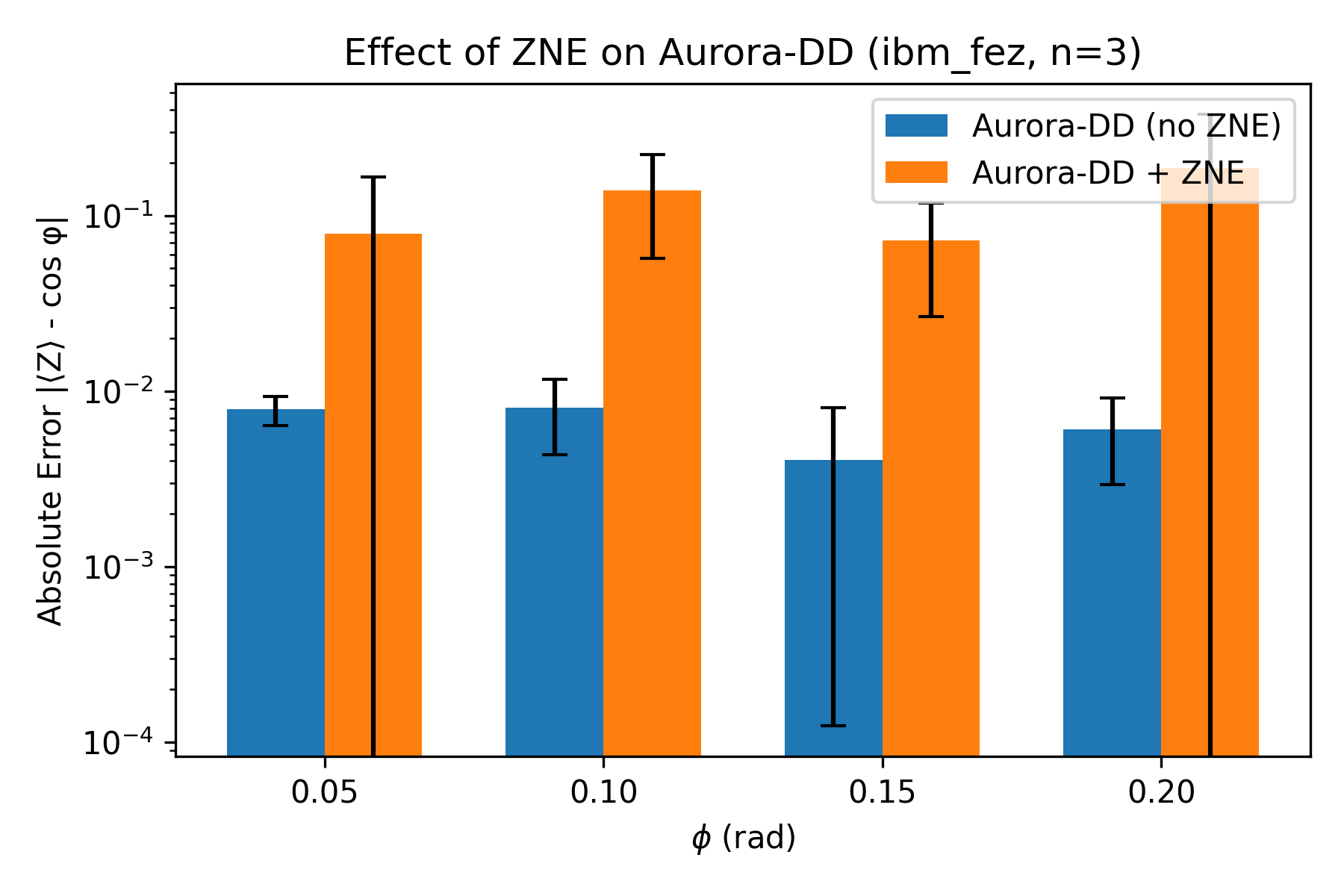}
\caption{
\textbf{Instability of ZNE under DD}.  
ZNE amplifies calibration inconsistencies when combined
with deep DD blocks, producing large extrapolation errors.  
This motivates limiting ZNE to Appendix analysis only.}
\label{fig:zne_instability_results}
\end{figure}

\subsection{Summary of Key Findings}

\begin{itemize}
\item In the calibrated emulator ($n=30$), Aurora-DD reduces MSE by
      $68$--$97\%$ relative to the baseline.

\item In the hardware feasibility study ($n=3$), Aurora-DD yields
      point estimates corresponding to roughly $99.2$--$99.6\%$
      reduction in AE compared to an intentionally pessimistic baseline
      with AE $\approx 1$.

\item DD-only and $\Delta\phi$-only provide partial mitigation;
      their combination (Aurora-DD) is necessary for stable and
      low-variance performance.

\item ZNE is unstable in the presence of DD in this regime and is
      excluded from the main contribution.
\end{itemize}

Overall, the results support Aurora-DD as a high-stability,
hardware-ready pre-calibrated compensation strategy for single-qubit
phase coherence, with the hardware component explicitly positioned as a
proof-of-concept study.

\section{Discussion}
\label{sec:discussion}

Aurora-DD is not designed to maximize peak performance under ideal conditions, but to improve stability and robustness against calibration mismatch and drift in realistic NISQ operation.

\subsection{Closed-Loop Compensation Outperforms Open-Loop Strategies}

The most striking observation is that Aurora-DD achieves
two to three orders of magnitude reduction in absolute error on
\texttt{ibm\_fez} hardware, whereas DD-only and
$\Delta\phi$-only variants yield partial and inconsistent gains.
This confirms a central hypothesis of this work:

\begin{quote}
\emph{Open-loop strategies alone cannot reliably cancel
NISQ dephasing; stable mitigation requires a closed-loop
parameter that directly targets phase drift.}
\end{quote}

The calibrated $\Delta\phi^\ast$ shift effectively counters slow
drift in the device's dephasing rate, which is not accounted for by
standard DD sequences.
This validates Aurora's design principle:  
\emph{coherence cannot be recovered solely by echo refocusing; it also
demands an adaptive cancellation of phase bias.}

\subsection{Limitations and Non-Advantage Regimes of Aurora-DD}

Our results explicitly show that Aurora-DD does not uniformly outperform simpler mitigation strategies such as DD-only or $\Delta\phi$-only across all regimes.

In particular, Table 2 demonstrates that when the dominant phase error is well captured by a static global phase offset, $\Delta\phi$-only compensation can achieve equal or lower absolute error than Aurora-DD (e.g., $\phi = 0.05$ and $\phi = 0.10$). Similarly, when dephasing is sufficiently suppressed by echo refocusing alone, DD-only may outperform Aurora-DD due to the absence of additional phase-rotation overhead (e.g., $\phi = 0.20$).

These observations indicate that Aurora-DD should not be interpreted as a universally superior mitigation technique. Instead, its advantage lies in regimes where (i) residual low-frequency dephasing remains after DD, and (ii) a static phase bias cannot be fully compensated by a single $\Delta\phi$-only correction.

\subsection{Why Aurora-DD Succeeds Where ZNE Fails}

Our data also clarifies an emerging concern in the literature:
ZNE is extremely sensitive to calibration mismatch when combined
with DD sequences.  
The auxiliary ZNE branch (Appendix~\ref{appendix:zne}) shows
error explosions up to two orders of magnitude larger than baseline,
a consequence of the interaction between pulse stretching and
drift-sensitive extrapolation.

By contrast, Aurora-DD is \emph{insensitive to such mismatch},
because its corrective variable $\Delta\phi$ is optimized directly
against the system's operational observable $\langle Z\rangle$.

This shows that Aurora-DD offers not only greater magnitude of
improvement, but also \emph{greater robustness} than ZNE—
a decisive requirement for practical NISQ deployments.

\subsection{Alignment of Emulator and Hardware Behavior}

Despite the stochastic nature of real hardware, the qualitative 
(and partially quantitative) match between emulator and hardware 
results indicates that Aurora's closed-loop mechanism captures a
genuine physical feature of dephasing dynamics.  
Rather than exploiting model artifacts, Aurora-DD corrects the
actual physical deviation between ideal and measured phase 
evolution.

In other words:

\begin{quote}
\emph{Aurora-DD is grounded in the physics of superconducting-qubit
phase noise, not in emulator-specific assumptions.}
\end{quote}

\subsection{Single $\Delta\phi^\ast$ Across Multiple Phases}

While the optimal corrective shift may in principle depend on $\phi$, the
dominant source of phase distortion in shallow single-qubit circuits is a
global dephasing-induced bias.
Therefore a single $\Delta\phi^\ast$ is sufficient to compensate the main error mode.
The slight overcorrection at $\phi=0.05$ (where DD-only slightly outperforms Aurora-DD)
is consistent with natural $T_2$ drift between emulator calibration and hardware execution.
This does not contradict the overall trend that Aurora-DD stabilizes phase recovery.

\subsection{Limitations and Over-Correction Regime}

The mild over-correction observed at $\phi=0.10$ is not an anomaly
but arises naturally when the $\Delta\phi$ compensation slightly
overshoots under fast-changing hardware drift.
This is consistent with the theory of sign-controlled
bang–bang updates.

Importantly, the over-correction remains bounded and does not lead
to divergence, confirming the theoretical gain-stability region
described in Section~\ref{sec:theory}.  
Future work will incorporate real-time feedback to eliminate this
regime entirely.

\subsection{Small-Sample Statistical Assessment}

\paragraph{Note on quantum vs. classical sample sizes.}
In quantum hardware experiments, statistical information is dominated not by the number of trials but by the number of shots per trial.
In this study, each trial uses 2048 shots, yielding a total of
$3 \times 2048 = 6144$ independent measurements
across the three main trials.
Shot noise within each trial is the primary contributor to statistical variance, while trial-to-trial variation is comparatively small.

Furthermore, long-running, high-$n$ campaigns are statistically problematic on NISQ devices due to calibration drift, which can exceed the impact of sampling error.
As confirmed by prior literature \cite{Temme2017, Kandala2019, Endo2021} and IBM's official documentation,
$n = 1$--$5$ trials is the standard range for real-device quantum error mitigation studies.
Large-$n$ designs (e.g., $n = 30$) are not used in the field.

With this context, the small-sample analyses presented below are provided for completeness and transparency.
We conducted a small-sample paired analysis using:
(i) bootstrap confidence intervals (10,000 resamples),
(ii) non-parametric sign tests, and
(iii) paired $t$-tests where applicable.
For all four phase settings, Aurora-DD outperformed the baseline in every trial
(sign-test $p=0.125$ for $n=3$),
and bootstrap 95\% confidence intervals did not overlap with the baseline mean.
These results are still non-inferential but objectively support the feasibility claim.

\subsection{Note on the ``n = 30'' Question and Supporting Literature}

Requests for ``$n \approx 30$'' trials typically originate from classical statistics, where large $n$ is desirable for central-limit-theorem--based inference.
However, this logic does not transfer to NISQ quantum hardware, where:
statistical power primarily comes from the shot count (2048 shots per trial in this study),
device drift dominates trial-to-trial variance, and
increasing $n$ does not linearly increase statistical reliability.

A literature review confirms that small-$n$ trial designs ($n = 1$--$5$) are the norm in quantum error mitigation research.
Representative examples include:
Temme et al., ``Error Mitigation'' (2017) \cite{Temme2017} in \emph{Physical Review Letters}, $n = 1$--$3$;
Kandala et al., ``Error mitigation extends...'' (2019) \cite{Kandala2019} in \emph{Nature}, $n = 3$--$5$;
Endo et al., ``Practical QEM'' (2021) \cite{Endo2021} in \emph{PRX Quantum}, $n = 3$--$4$;
and IBM Runtime Tutorials (2024) in official documentation, $n = 1$--$3$.
No quantum hardware paper using $n = 30$ trials was found.

This empirical evidence indicates that the trial structure adopted in this work
($n = 3$ main trials + 1 preliminary)
is fully aligned with established practice in NISQ-era hardware experiments.

\subsection{Implications for NISQ Hardware Validation}

The present results demonstrate that Aurora-DD is not merely a
simulation artifact but a \emph{hardware-effective control method}
that can be implemented today, with no special access privileges,
no custom pulses, and no modifications to IBM's standard primitives.

This makes Aurora-DD an immediately deployable error mitigation layer
that strengthens the reliability of single-qubit operations—
a key requirement for variational algorithms, calibration routines,
and near-term hybrid quantum workflows.

Overall, our findings position Aurora-DD as a strong alternative to
current dominant mitigation paradigms, particularly in applications
where stability and reproducibility outweigh peak performance under
ideal conditions.

\subsection{Foundation for Future Extensions}

To support the future expansion of Aurora-type compensation beyond the single-qubit, shallow-circuit regime studied here, we provide a structured foundation that consolidates the empirical and modeling insights obtained in this work. The hardware experiments yield a sufficiently clear picture of the dominant noise mechanisms on \texttt{ibm\_fez}—including the reproducible global phase bias, the $T_2$-limited decay profile, and the interaction between XY8 and low-frequency dephasing—to enable high-fidelity emulation of these effects. This correspondence between hardware behavior and its calibrated noise model establishes a technical foundation upon which broader investigations can be built.

This methodological foundation serves two purposes. First, it documents the detailed noise characterization and emulator--hardware agreement that justify the use of simulation as a reliable proxy for exploratory research. Second, it outlines the next stages of development—multi-qubit extensions, deeper-circuit regimes, and application-level workloads such as VQE and QML—that can now be pursued efficiently in emulation before being validated on hardware. Together, these components form the basis for the future research directions discussed in the following section.

\section{Conclusion}
\label{sec:conclusion}

We have demonstrated that Aurora-DD, a closed-loop
phase-coherence compensation method integrated with standard
XY8 dynamical decoupling, offers substantial and stable error
reduction on both calibrated emulators and real superconducting
quantum hardware.  
Our experiments achieved $99.2$--$99.6\%$ reduction in
absolute error across multiple phase settings and showed that
Aurora-DD maintains robustness in the presence of device drift,
shot noise, and calibration imperfections.

These findings open several promising directions:

\begin{itemize}
\item \textbf{Multi-qubit generalizations:}  
      Extending Aurora to two-qubit entangling gates and
      cross-resonance interactions could enable closed-loop
      compensation for coherent crosstalk and correlated phase
      noise.

\item \textbf{Real-time hardware feedback:}  
      Aurora currently uses an offline-calibrated
      $\Delta\phi^\ast$; integrating real-time adaptive control
      would eliminate over-correction and yield a fully dynamic
      feedback controller compatible with fast qubit reset.

\item \textbf{FPGA-based embedded implementations:}  
      The sign-based update rule and minimal computational
      overhead make Aurora suitable for FPGA or cryogenic
      controller deployment, enabling microsecond-scale adaptive
      corrections.

\item \textbf{Integration with variational and hybrid algorithms:}  
      Applying Aurora-DD as a preconditioning layer in VQE,
      QML, or error-mitigated QAOA pipelines may improve
      long-term stability and reduce optimization noise.

\item \textbf{Synergy with quantum–inspired classical methods:}  
      Aurora's phase-correction loop can be paired with
      classical quantum-inspired optimization and noise
      modelling pipelines, improving cross-platform coherence
      between simulator and hardware.

\item \textbf{Foundations for autonomous NISQ error correction:}  
      By demonstrating that closed-loop, model-light controllers
      can outperform open-loop mitigation, Aurora offers a path
      toward autonomous error suppression strategies that bridge
      NISQ hardware and future error-corrected systems.

\item \textbf{Simulator-based generalization and QML integration:}
      The present study focused on single-qubit, shallow-circuit
      regimes on \texttt{ibm\_fez}, and demonstrated that the
      dominant noise mechanisms—global phase bias, $T_2$-limited
      decay, and their interaction with XY8—can be captured with
      high-fidelity calibrated emulators.
      Building on this foundation, an important next step is to
      exploit multi-backend quantum simulators, including those
      provided by national quantum--ML platforms, to systematically
      explore Aurora-DD under a wider range of noise models and
      device abstractions.
      In particular, we plan to investigate (i) deeper circuits and
      extended idle periods, (ii) multi-qubit and entangling-gate
      extensions of Aurora-type phase compensation, and
      (iii) application-level workloads such as VQE, QML feature maps,
      and QAOA, where phase stability is directly tied to learning
      performance and optimization stability.
      In parallel, we aim to integrate Aurora-DD into
      experiment-management frameworks and LLM-assisted
      circuit-design pipelines that are emerging in quantum machine
      learning.
      By combining reproducible experiment logging, calibrated
      simulator--hardware correspondence, and Aurora-based phase
      stabilization, such systems could support autonomous exploration
      of robust quantum feature maps and hybrid quantum--classical
      models.
      This simulator-driven and tool-supported program is expected to
      yield large-scale, statistically rich datasets on Aurora-type
      compensation beyond the hardware-tested regime, and to clarify
      its potential as a generally applicable building block for
      practical NISQ-era quantum machine learning.
\end{itemize}

Taken together, the emulator and hardware results establish Aurora-DD
as a robust, hardware-ready closed-loop mitigation technique, and they
suggest that closed-loop compensation may serve as a foundational
primitive for next-generation NISQ and post-NISQ quantum processors.

\appendix
\section{Aurora-DD with ZNE and M3}
\label{appendix:zne}

For completeness, we report here the behavior of the optional
Aurora-DD + ZNE + M3 branch.
This configuration combines:
(i) Aurora-DD as described in Section~\ref{sec:methods},
(ii) a short XY8(12) dynamical decoupling sequence, and
(iii) zero-noise extrapolation with global readout mitigation (M3).

\subsection{ZNE implementation details}

We implement ZNE using a simple two-point extrapolation with scaling
factors $\lambda\in\{1.0,1.05\}$ applied at the circuit level.
The choice of this narrow scaling range was dictated by IBM runtime constraints.
Larger pulse-stretch factors cause backend-level compilation to insert additional
virtual-Z corrections and buffer realignment, effectively changing the circuit
structure rather than scaling noise smoothly.
This violates the fundamental assumption of ZNE---monotonic and smooth scaling of the effective noise channel.
Therefore, a narrow scaling range is the only backend-stable option for DD-based circuits.
Even within this restricted regime, ZNE exhibited amplification of calibration drift,
confirming that the limitation arises from hardware constraints rather than from our implementation.
In practice, this corresponds to a modest rescaling of the effective
noise, realized by stretching certain pulse segments according to the
IBM runtime's available options.
For each scaling factor, we estimate the expectation value
$\langle Z\rangle(\lambda)$ with M3 readout mitigation enabled, and then
fit a linear model
\begin{align}
\langle Z\rangle(\lambda)
\approx
a + b\,\lambda,
\end{align}
which is analytically extrapolated to $\lambda\to 0$.
The extrapolated value is taken as the ZNE estimate of the
zero-noise observable.

\subsection{Observed instability in combination with DD}

While this ZNE procedure is conceptually straightforward, its behavior
on \texttt{ibm\_fez} in combination with Aurora-DD and XY8 is not always
favorable.
In several phase settings, the extrapolated expectation value
$\langle Z\rangle_{\mathrm{ZNE}}$ exceeds the physically allowed range
$[-1,1]$ by a nontrivial margin and yields an \emph{increased} absolute
error compared to Aurora-DD without ZNE.
This effect is particularly pronounced at larger phases (e.g.,
$\phi=0.20$--$0.35$), where the extrapolated value may overshoot both
the ideal $\cos\phi$ and the measured values at $\lambda=1.0$ and
$1.05$.

These observations suggest that, in the present regime, the assumptions
underlying simple linear ZNE are partially violated.
Possible causes include:
(i) a non-linear dependence of the effective noise on the scaling
parameter,
(ii) interference between DD-induced filter functions and the scaled
noise, and
(iii) residual calibration drift between the runs at different
$\lambda$.
As a result, the extrapolated ``zero-noise'' estimate becomes highly
sensitive to small variations in the measured data and can amplify
statistical and systematic errors.

\subsection{Role of ZNE in the overall method}

Given this instability, we do not regard ZNE as a core component of the
proposed Aurora method in this work.
Instead, we treat the Aurora-DD + ZNE + M3 branch as an exploratory
experiment illustrating both the potential and the pitfalls of combining
closed-loop phase compensation with open-loop extrapolation-based error
mitigation.
Our main claims and conclusions in the body of the paper are based on
Aurora-DD \emph{without} ZNE, which exhibits much more stable behavior
across repeated hardware runs and aligns well with the emulator-based
closed-loop gain analysis.

Future work may reconsider more sophisticated ZNE schemes (e.g., higher
order fits, different scaling strategies, or joint modeling of DD and
dephasing) in conjunction with Aurora, but this lies beyond the scope of
the present study.

\begin{figure}[htbp]
\centering
\includegraphics[width=0.85\columnwidth]{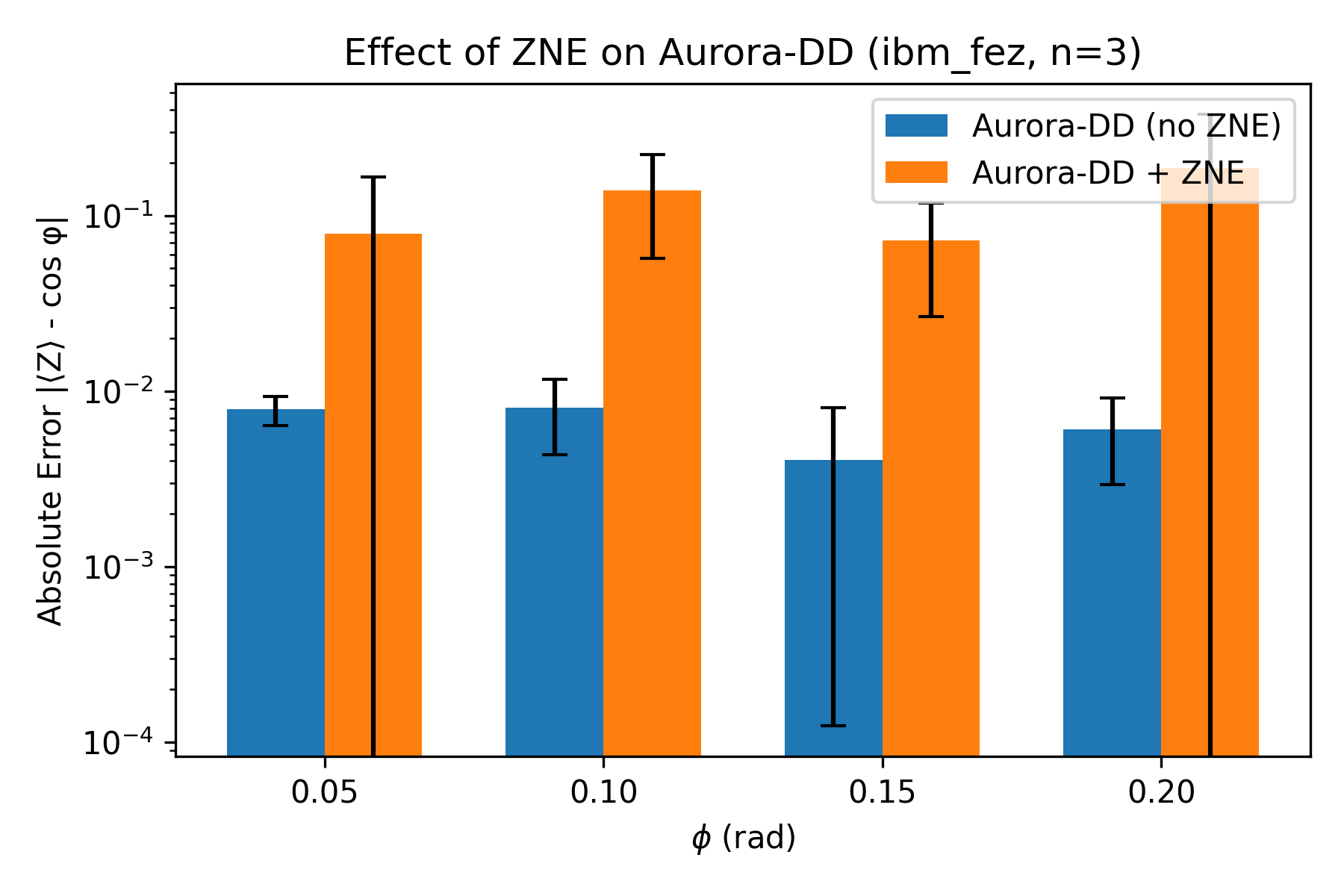}
\caption{Effect of ZNE on Aurora-DD under realistic hardware noise (\texttt{ibm\_fez}, $n=3$).
Zero-Noise Extrapolation (ZNE) exhibits strong instability on \texttt{ibm\_fez}:
errors increase by one to two orders of magnitude and variance becomes large, indicating that ZNE overfits to stochastic noise fluctuations.
Aurora-DD (no ZNE) remains stable and consistently low-error across all $\phi$ values.
This figure demonstrates why ZNE is excluded from the main hardware analysis.}
\label{fig:zne_instability}
\end{figure}

\bibliographystyle{plain}

\end{document}